\documentclass[runningheads]{llncs}
\usepackage[T1]{fontenc}

\usepackage{amsmath,amsfonts}
\usepackage{algorithmic}
\usepackage{textcomp}
\usepackage{xcolor}
\usepackage{listings}
\usepackage{colortbl}
\usepackage{etoolbox}
\usepackage{graphicx}
\usepackage{xspace}
\usepackage{url}
\usepackage{booktabs}
\usepackage{multicol}
\usepackage{tikz}
\usetikzlibrary{arrows.meta, positioning}
\usepackage{eso-pic}

\def\BibTeX{{\rm B\kern-.05em{\sc i\kern-.025em b}\kern-.08em
    T\kern-.1667em\lower.7ex\hbox{E}\kern-.125emX}}

\lstdefinelanguage{Julia}%
  {morekeywords={abstract,break,case,catch,const,continue,do,else,elseif,end,export,false,for,function,struct,immutable,import,importall,if,in,macro,module,otherwise,quote,return,switch,true,try,type,typealias,using,while},%
   morekeywords=[2]{AbstractDiscreteTimeStochasticSystem, AdditiveGaussianLinearSystem, AdditiveGaussianPolySystem, AdditiveGaussianUncertainPWASystem, synthesize_barrier, sos_system_specific_constraints, ConstantBarrier, SOSBarrier, zeros, ones,
   transition_probabilities, load_probabilities,
   open_dataset, generate_partitions,
   load_dynamics,
   SumOfSquaresBarrierAlgorithm, ConstantBarrierAlgorithm,
   SumOfSquaresAlgorithm(), SumOfSquaresAlgorithm,
   DualAlgorithm(), DualAlgorithm,
   IterativeUpperBoundAlgorithm(), IterativeUpperBoundAlgorithm,
   GradientDescentAlgorithm(), GradientDescentAlgorithm, CEGISAlgorithm
   },  
   morekeywords=[3]{Hyperrectangle, @polyvar, UnionSet}, 
   sensitive=true,%
   morecomment=[l]\#,%
   morecomment=[n]{\#=}{=\#},%
   morestring=[s]{"}{"},%
   morestring=[m]{'}{'},%
}[keywords,comments,strings]%

\usepackage[numbers]{natbib}

\RequirePackage{graphicx}
\usepackage[firstpageonly=true,angle=0,vpos=.147\paperheight,hpos=.39\linewidth,vanchor=t,hanchor=l]{draftwatermark}
\SetWatermarkText{%
\raisebox{-4cm}
{\begin{minipage}{\linewidth}\centering\includegraphics[width=12.5mm]
{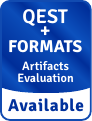}\hfill\includegraphics[width=12.5mm]
{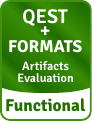}\end{minipage}}%
}

\newif\ifarxiv
\arxivtrue

\usepackage{hyperref}
\newcommand{\orcid}[1]{\href{https://orcid.org/#1}{\includegraphics[width=3mm]{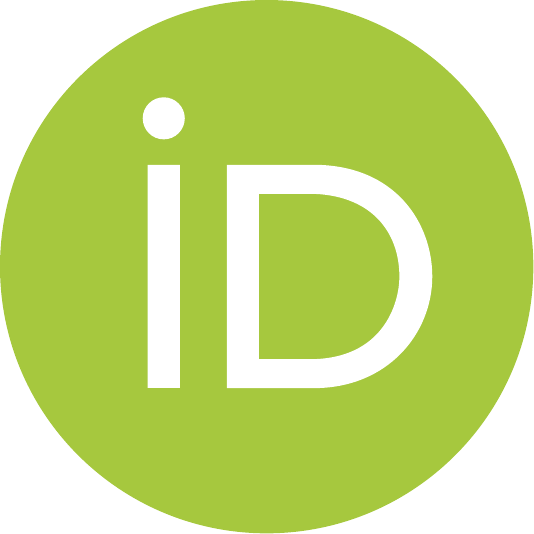}}}

\begin{document}
\AddToShipoutPictureBG*{%
  \AtPageUpperLeft{%
    \hspace{19cm}%
    \raisebox{-1.5cm}{%
      \makebox[0pt][r]{%
        \shortstack[r]{%
          To appear in the Third International Joint Conference on Quantitative Evaluation of Systems\\
           and Formal Modeling and Analysis of Timed Systems (QEST+FORMATS 2026)%
        }%
      }%
    }%
  }%
}
\title{StochasticBarrier.jl: A Toolbox for Stochastic Barrier Function Synthesis}
%
%
\author{Rayan Mazouz\inst{1}
\orcid{0009-0002-8440-2551} 
\and
Frederik Baymler Mathiesen \inst{2}
\orcid{0000-0002-2243-0445} 
\and
Luca Laurenti \inst{2}
\orcid{0000-0003-1190-6097} 
\and\\
Morteza Lahijanian \inst{1}
\orcid{0000-0001-7549-4365}
}
\authorrunning{Mazouz et al.}
%
\institute{University of Colorado Boulder, Boulder CO, United States \and
Delft University of Technology, The Netherlands
}

\newcommand{\RM}[1]{{\color{cyan}[RM: #1]}}
\newcommand{\JS}[1]{{\color{purple}[JS: #1]}}
\newcommand{\ml}[1]{{\color{blue}[ML: #1]}}
\newcommand{\LL}[1]{{\color{teal}[LL: #1]}}

\newcommand{\pX}{\mathbf{x}}
\newcommand{\pV}{\mathbf{v}}
\newcommand{\pW}{\mathbf{w}}
\newcommand{\pU}{\mathbf{u}}
\newcommand{\NN}{f^w}
\newcommand{\cov}{R}
\newcommand{\erf }{\text{erf}}
\newcommand{\dyn}{f(\pX_{k}, \pU_{k}) }
\newcommand{\inv}{A_{\text{inv}}}

\newcommand{\up}[1]{\overline{#1}}
\newcommand{\low}[1]{\underline{#1}}

\newcommand\stateset{X}
\newcommand\controlset{U}
\newcommand{\unsafe}{\mathrm{u}}
\newcommand{\safe}{\mathrm{s}}
\newcommand\safethresh{p^*}
\newcommand{\initial}{\mathrm{0}}
\newcommand{\set}[1]{\{#1\}}
\newcommand{\ctrl}{\mathrm{c}}
\newcommand{\ControlSafeSet}{C_\safe^{N}}
\newcommand{\ProbControlSafeSet}[1]{C_{\safethresh}^{N,#1}}

\newcommand{\mean}{\mathbf{m}}
\newcommand{\Gm}{\mean(\mathbf{x}_{k})}
\newcommand{\Gv}{{Q}(\mathbf{x}_{k})}
\newcommand{\lGm}{\low{\mean}(\mathbf{x}_{k})}
\newcommand{\uGm}{\up{\mean}(\mathbf{x}_{k})}
\newcommand{\kernel}{\kappa}
\newcommand{\std}{\sigma}
\newcommand{\var}{\std^2}
\newcommand{\GInput}{X}
\newcommand{\dataset}{D}
\newcommand{\datasetsize}{M}

\newcommand{\indicator}[1]{\mathbbm{1}_{#1}}
\newcommand{\B}{\mathcal{B}}
\newcommand{\expect}{\mathbb{E}}
\newcommand{\pv}{\mathbf{v}}
\newcommand{\pw}{\mathbf{w}}
\newcommand{\px}{\mathbf{x}}
\newcommand{\pz}{\mathbf{z}}

\newcommand{\reals}{\mathbb{R}}
\newcommand{\naturals}{\mathbb{N}}
\newcommand{\naturalszero}{\mathbb{N}_{\geq 0}}
\newcommand{\complex}{\mathbb{C}}
\newcommand{\N}{\mathcal{N}}
\renewcommand{\L}{\mathcal{L}}
\newcommand{\pdf}{p}
\newcommand{\pdfx}{\pdf_{\pX}}


\newcommand{\tempop}[1]{\mathcal{#1}}
\newcommand{\until}{\,\tempop{U}}
\newcommand{\eventually}{\,\Diamond} 
\newcommand{\globally}{\,\Box} 
\renewcommand{\diamond}{\Diamond}

\newcommand{\pstlnn}{PrSTL$_{nn}$\xspace}
\newcommand{\stlnn}{STL$_{nn}$\xspace}
\newcommand{\timedomain}{\mathbb{T}}
\newcommand{\TA}{\mathcal{T} \hspace{-1mm} \mathcal{A}}

\newcommand{\refhere}{{\color{red} ref here}}

\definecolor{darkblue}{rgb}{0.0, 0.0, 0.5}  
\definecolor{darkmagenta}{rgb}{0.5, 0, 0.5}  
\definecolor{darkgreen}{rgb}{0, 0.4, 0}  

\newcommand{\juliafile}[1]{{\color{darkblue}{\small\texttt{#1}}}}
\newcommand{\juliafunction}[1]{{\color{darkmagenta}{\small\texttt{#1}}}}

\newcommand{\tool}{\texttt{StochasticBarrier.jl}\xspace}

\newcommand{\rv}[1]{\textcolor{blue}{#1}}
\newcommand{\del}[1]{\textcolor{red}{#1}}
\maketitle              
\begin{abstract}
We present \tool, an open-source Julia-based toolbox for generating Stochastic Barrier Functions (SBFs) for safety verification of discrete-time stochastic systems with additive Gaussian noise. \tool certifies linear, polynomial, and piecewise affine (PWA) systems.  The latter enables verification for a wide range of system dynamics, including general nonlinear types. The toolbox implements a Sum-of-Squares (SOS) optimization approach, as well as methods based on piecewise constant (PWC) functions. For SOS-based SBFs, \tool leverages semi-definite programming solvers, while for PWC SBFs, it offers three engines: two using linear programming (LP) and one based on gradient descent (GD). Benchmarking \tool against the state-of-the-art shows that the tool outperforms existing tools in computation time, safety probability bounds, and scalability across over 30 case studies. Compared to its closest competitor, \tool is up to four orders of magnitude faster, achieves significant safety probability improvements, and supports higher-dimensional systems.
\end{abstract}

\section{Introduction}
\label{sec:intro}

Stochastic barrier functions (SBFs) are a class of tools used for the formal verification of stochastic systems \cite{prajna2007framework, SANTOYO2021109439}. These functions establish a lower bound on the probability of a system remaining within a safe set, using martingale theory \cite{SANTOYO2021109439, mazouz2022safety}. A key advantage of SBFs over other verification methods is their capacity to analyze the time evolution of a system through a set of static constraints, eliminating the need to propagate system uncertainty. Instead, SBF synthesis methods solve a functional optimization problem, for which various approaches have been developed based on, e.g., Sum-of-Squares (SOS) polynomials \cite{prajna2007framework, SANTOYO2021109439, mazouz2022safety} and piecewise (PW) functions \cite{mazouz2024piecewise, mazouz2024data, mathiesen2024data}. Despite their potential, however, there has been limited effort to develop robust and efficient tools for SBF synthesis, restricting their broader application in practical fields such as robotics and autonomous vehicles.

In this paper, we address this gap by introducing
an open-source Julia-based toolbox that generates SBFs for \emph{discrete-time stochastic systems} with \emph{additive Gaussian noise} - \tool
\ifarxiv
    \footnote{
    \url{https://github.com/aria-systems-group/StochasticBarrier.jl}.}.
\else
    \footnote{
    \url{https://github.com/HSCC2026/StochasticBarrier.jl}.}.
\fi
The toolbox supports a wide range of dynamics, namely, \emph{linear}, \emph{polynomial}, and \emph{PW affine (PWA) inclusions}, the latter enabling the verification of non-linear, non-polynomial systems. The established SOS optimization method is implemented,  
    as well as more recent approaches based on PW Constant (PWC) functions. For SOS optimization, existing semi-definite programming (SDP) solvers are leveraged. For PWC-based SBFs, three engines are included: two that utilize Linear Programming (LP) and one based on Gradient Descent (GD), as introduced in \cite{mazouz2024piecewise}.

The toolbox is developed in Julia, a modern programming language designed for the scientific community. 
Julia combines the interactivity and dynamic typing of scripting languages, facilitating fast prototyping, with the high performance of compiled languages \cite{perkel2019julia}. Also, Julia's support for parametric typing allows for customizable precision in numerical computations~\cite{mathiesen2024intervalmdp}. 
We evaluate \tool on various benchmarks and compare its performance against \texttt{PRoTECT} \cite{wooding2024protect} and \texttt{StochasticBarrierFunctions}~\cite{SANTOYO2021109439}, two
state-of-the-art tools, 
which, to the best of our knowledge, are the only available SBF tools for discrete-time stochastic systems. These existing tools exclusively support SOS-based approaches. In addition, the three PWC-SBF methods are compared against the SOS-method.
Benchmarks include over 30 case studies across eight different systems with linear, polynomial, and PWA inclusion dynamics, ranging from \emph{one} to \emph{six} dimensions. The results demonstrate that \tool outperforms existing tools in three metrics: computation time, safety probability bounds, and scalability. 
Compared to its closest competitor, \tool is up to $1000\times$ faster for successful synthesis, improves safety probability bounds from $0$ to $1.0$ in some cases, and handles systems with twice the dimensionality.

\subsection{Related Work}
Barrier certificates are synthesized to prove safety for dynamical systems in both deterministic \cite{schneeberger2023sos, wang2018permissive} and stochastic settings \cite{prajna2002introducing, jagtap2020formal}. Stochastic barrier functions (SBFs) are used to obtain a formal lower bound on the probability of remaining safe, starting from an initial set, under a well-defined probability measure. SBFs have been applied to verify safety for linear and polynomial systems \cite{SANTOYO2021109439}, nonlinear systems \cite{mazouz2022safety}, and data-driven systems \cite{mazouz2024data}.

Sum-of-Squares (SOS) optimization is a popular method for SBF synthesis \cite{prajna2002introducing, prajna2007framework, SANTOYO2021109439, jagtap2020formal, mazouz2022safety}. Some works utilize SOSTOOLS \cite{prajna2002introducing} to obtain these SBFs. The main downside of SOSTOOLS-based SBFs is that the algorithms are embedded in MATLAB, which scales poorly when bridging to SDP.  PRoTect \cite{wooding2024protect}, a Python based implementation, scales better, but is limited to low-degree SOS barriers and systems of polynomial class. Our toolbox, which is implemented in Julia, provides significant speedups for intensive numerical optimization tasks and memory management \cite{perkel2019julia, coleman2021matlab}. 

Related tools for the verification of stochastic systems in Julia include \emph{IntervalMDP.jl} \cite{mathiesen2024intervalmdp}, which performs CUDA-accelerated 
value iteration for Interval Markov Decision Processes (IMDPs). JuliaReach \cite{bogomolov2019juliareach} performs set-based reachability for deterministic systems.
A tool for Satisfiability modulo theories, Satisfiability.jl 
\cite{soroka2023satisfiability}, provides high-level representations for SMT formulae. Finally, ModelVerification.jl
\cite{wei2024modelverification} verifies DNNs and safety specifications. None of these verify safety for stochastic systems using SBFs. 
A platform for this purpose is provided in \tool.

The remainder of the paper is organized as follows. In Section \ref{sec:toolbox}, we provide a detailed overview of the toolbox, with instructions and examples on setting up system models and invoking SBF synthesis engines. 
In Section \ref{sec:studies}, we demonstrate the performance of \tool through a series of case studies and benchmark comparisons.
Finally, we conclude the paper with a summary and closing remarks in Section~\ref{sec:conclusion}.

\section{Theoretical Framework}
\label{sec:theory}
We provide an overview of the SBF theoretical framework. An outline of the problem formulation is provided, followed by a review of the implemented methods. For more detail, we refer the reader to \cite{mazouz2022safety,mazouz2024piecewise}.
\subsection{Problem} 
\noindent Consider a stochastic process defined by difference equation
\begin{equation}
    \label{eq:system}
    \px_{k+1} = f(\px_k) + \pw_k,
\end{equation}
where $ \px_{k} \in \mathbb{R}^n$ is the state, 
$f:\mathbb{R}^n \to \mathbb{R}^n$ is the vector field representing the one-step dynamics of System~\eqref{eq:system}, and
$\pw_k \in \mathbb{R}^n$ is an independent and identically distributed random variable, representing noise.
The probability distribution of $\pw_k$ is assumed to be zero-mean Gaussian, i.e., $\pw_k \sim p_{\pw} =\mathcal{N}(0, \Sigma)$, where $\Sigma = \mathrm{diag}\!\left(\sigma_1^2,\ldots,\sigma_n^2\right)$.
We define the stochastic kernel 
$T(X\mid x):= \int_{\mathbb{R}^{\mathrm{w}}} \mathbf{1}_X
(f(x) + w)p_{\mathbf{w}}(dw)$.
It follows from this definition of $T$ that, given an initial condition $\mathbf{x}_0 = x_0 \in \reals^{n}$, $\mathbf{x}_k$ is a Markov process \cite{bertsekas2004stochastic}. 
 Let $X_\safe \subset \mathbb{R}^n$ be a bounded set representing the safe set, $X_0 \subseteq X_\safe$ be the initial set, and $N \in \mathbb{N}$ be the time horizon. Then, \emph{probabilistic safety} is defined as
$P_{\safe}(X_\safe,X_0,N) = \inf_{x_0 \in X_0} \mathrm{Pr}^{x_0}[\forall k \in \{0,...,N\}, \pX_k \in X_\safe].$ 

To reason about $P_{\safe}(X_\safe,X_0,N)$, {\textit{Stochastic Barrier Function} (SBFs) can be used \cite{SANTOYO2021109439, mazouz2022safety}. 
Function $B:\mathbb{R}^n \to \mathbb{R}_{\geq 0}$ is a SBF for System~\eqref{eq:system} if there exist scalars $\eta,\beta \geq 0$, such that
\begin{subequations}
    \begin{align}
        &B(x) \geq 1 \qquad &&\forall x\in \mathbb{R}^n\setminus X_s\label{eq:barrier_unsafe},\\
        &B(x) \leq \eta \qquad &&\forall x\in X_0\label{eq:barrier_initial},\\
        &\mathbb{E}[B(f(x) + \pw))] \leq  B(x) + \beta \qquad && \forall x\in X_s \label{eq:barrier_expectation},
    \end{align}
\end{subequations}
which guarantees a lower-bound for the \emph{probabilistic safety} 
\begin{equation}
    \label{eq:prob}
    P_{\safe}(X_\safe,X_0,N) \geq 1-( \eta + N \beta ).
\end{equation}
The goal of the SBF synthesis problem is to find a $B(x)$ that maximizes $P_s$, subject to Conditions \eqref{eq:barrier_unsafe}-\eqref{eq:barrier_expectation}.
\tool solves this problem by computing a $B$ along with a $P_s$ lower bound.

From a computational perspective, solving this functional optimization problem is non-trivial. Several methods have been developed to ensure the convexity of the problem.
One well-established approach assumes $B$ is a polynomial of a given degree, which allows the problem to be formulated as a SOS optimization, ultimately leading to a semi-definite program~\cite{SANTOYO2021109439, prajna2006barrier, mazouz2022safety}.
A more recent approach, introduced in~\cite{mazouz2024piecewise}, assumes $B$ is a piecewise constant (PWC) function, reducing the optimization problem to a linear program.
Both methods are implemented in \tool.

\subsection{Implemented SBF Methods}
\tool supports polynomial (SOS) and PWC class-es of SBFs for System~\eqref{eq:system}, where $f$ is affine, polynomial, or bounded by piecewise affine (PWA) functions. 

\subsubsection{Sum-of-Squares (SOS)}
A multivariate polynomial $\lambda(x)$ with $x \in \mathbb{R}^{n}$ is an SOS polynomial if there exist polynomials  $\lambda_{i}$, $i = 1, \ldots, r$, for ${r \in \naturals}$, such that   $\lambda(x) = \sum_{i=1}^{r} \lambda_{i}^{2}(x).$ If $\lambda(x)$ is SOS, then $\lambda(x) \geq 0$ for all $x \in \mathbb{R}^{n}$. The set of SOS polynomials is denoted by $\Lambda(x)$, and a vector of SOS polynomials by $\mathcal{L}(x)$. %
Under the assumption that $f$ in System~\eqref{eq:system} is a polynomial of degree one or higher, and let $X_0, X_s$ be compact semi-algebraic sets, then, the following theorem~\cite{SANTOYO2021109439} is implemented in \tool to reduce the SOS-SBF synthesis problem to a convex program.
\begin{theorem}[SOS-SBF for Polynomial Systems \cite{SANTOYO2021109439}]
    \label{th:SOS-poly}
    Let the safe set $ X_\safe = \set{  x \in \reals^{n} \mid h_{\safe}(x) \geq 0}$, 
    initial set $ X_{\initial} = \set{  x \in \reals^{n} \mid h_{\initial}(x) \geq 0}$, 
    and unsafe set $X_{\unsafe} = \reals^n \setminus X_\safe = \set{  x \in \reals^{n} \mid h_{\unsafe}(x) \geq 0}$,
    where $h_s,h_0,h_u$ are vectors of polynomials, be given.
    Then, a SBF $B(x) \in \Lambda(x)$ for System~\eqref{eq:system} is
    obtained by solving the following SOS optimization problem
    \begin{subequations}
    \label{eq: sos optimization}
    \begin{align}
        &\min_{\beta,\eta \geq 0} \quad \eta + N \beta  \qquad
         \text{subject to:} \label{eq: SOS optimization objective} \\
        & \hspace{4mm} B(x)\in \Lambda(x), \label{constraint1} \\
        &  \hspace{.5mm} - B(x) - \mathcal{L}^{T}_{\initial}(x)h_{\initial}(x) + \eta \in \Lambda(x), \label{constraint2}  \\
        & \hspace{4mm} B(x) - \mathcal{L}^{T}_{\unsafe}(x)h_{\unsafe}(x) -1 \in \Lambda(x), \label{constraint3} \\
       & \hspace{.5mm} -\mathbb{E}[B(f(x) + \pw)] + B(x) + \beta - \mathcal{L}^{T}_{s}(x)h_{s}(x) \in \Lambda(x), \label{constraint4_poly}
    \end{align}
    \end{subequations}
    where $\mathcal{L}_{0}$, $\mathcal{L}_{h}$, and $\mathcal{L}_{s}$ are vectors of SOS polynomials of appropriate dimension. 
    A valid solution to this optimization guarantees safety probability in Eq.~\eqref{eq:prob}.
\end{theorem}

\paragraph{PWA Inclusion Dynamics.}

Additionally to polynomial $f$, non-polynomial $f$ are also supported in \tool by supplying piecewise affine (PWA) bounds of $f$. In particular, given a partition $Q=\{q_1, \ldots ,q_{|Q|} \}$ of the safe set $X_\safe$ for every $q \in Q$, \tool supports $f$ expressed via bounds 
\begin{equation}
\label{eqn:UpperandLowerBound}
     \low{f}_q(x)= \low{A}_q x + \low{b}_q \quad \text{and} \quad \up{f}_q(x)= \up{A}_q x + \up{b}_q
\end{equation}
such that $\low{f}_q(x) \leq f(x) \leq \up{f}_q(x)$ for every $x \in q$. 
The following theorem shows a convex optimization framework for synthesis of SOS-SBF for such dynamics.

\begin{theorem}[SOS-SBF for PWA Inclusion Systems \cite{mazouz2022safety}]
    \label{th:SOS-general}
    Consider function $B(x)$ and sets $X_0$, $X_s$, $X_u$ as in Theorem~\ref{th:SOS-poly} and $X_s$-partition $Q = \{q_1, \ldots, q_{|Q|}\}$, where each $q \in Q$ is a semi-algebraic set $q= \set{  x \in \reals^{n} \mid h_{{q}}(x) \geq 0 }$.
    Given PWA functions $\low{f}_q$ and $\up{f}_q$ over $Q$ as defined in Eq.~\eqref{eqn:UpperandLowerBound}, 
    an SOS SBF $B(x)$ for System~\eqref{eq:system}, with general $f(x) \in [\low{f}_q(x), \up{f}_q(x)]$, can be obtained by solving 
    the SOS optimization problem in \eqref{eq: sos optimization} and replacing Constraint~\eqref{constraint4_poly} with 
    $-\mathbb{E}[B( y + {w}) \mid x]  + B(x) + \beta -  \mathcal{L}^{T}_{q,x}(x)h_{q}(x) - 
         \mathcal{L}^{T}_{q, y}(x) 
         \big(
         (\up{f}_q(x) - y)\odot(y - \low{f}_q(x))
         \big)\in \Lambda(x, y),$  for all $q \in Q$, 
    where $\odot$ is the Schur product. 
    \end{theorem}
In \tool, the implementation of partition $Q$ is based on a grid, where each $q \in Q$ is defined as a hyper-rectangle.
Then, provided partition $Q$ and PWA functions $\low{f}_q$ and $\up{f}_q$, the tool performs the above optimization. 

\subsubsection{PWC Method}
\label{subsec:pwcmethods}
\tool~ also supports the class of PWC-SBFs.
Consider a partition $Q = \{q_1, \ldots, q_{|Q|}\}$ of $X_s$
such that vector field $f(x)$ is continuous in each region $q \in Q$. 
Define PWC function $B: \mathbb{R}^n \to \mathbb{R}$ as
\begin{equation}
    \label{eq:pwf}
    B(x) = 
    b_i \; \text{ if } \; x \in q_i, \text{ otherwise } B(x) = 1.
\end{equation}
where $b_i \in \mathbb{R}$.
Further, let $q_u = X_u$ and denote the bounds on the transition kernel $T$ from partition $q_i\in Q$ by, for all $j \in \{1,\ldots, |Q|, u\}$,
\begin{equation}
    \label{eq:kernel bounds}
    \low{p}_{ij} = \inf_{x \in q_i} T(q_j \mid x), \quad \text{and} \quad \up{p}_{ij} = \sup_{x \in q_i} T(q_j \mid x).
\end{equation}
Then, the set of feasible values for $T$ from $q_i$ is given by
$\mathcal{P}_i = \big\{  p_i = (p_{i1},\ldots,p_{i|Q|},  p_{i\unsafe})  \in [0,1]^{|Q|+1}$ \; s.t. \;
$\sum_{j=1}^{|Q|} p_{ij}+p_{i\unsafe} = 1, 
        \; \low{p}_{ij} \leq p_{ij} \leq \up{p}_{ij} \;\;\; \forall j \in \{1,\ldots,|Q|,\unsafe \} \big \}.$ 
The following theorem formalizes the optimization problem for PWC class of SBFs for System~\eqref{eq:system}.
\begin{theorem}[PWC-SBF \cite{mazouz2024piecewise}]
    \label{th:pwa}
    Consider System~\eqref{eq:system} with safe set $X_s$, initial set $X_0$, and partition $Q = \{q_1,\ldots, q_{|Q|}\}$ of $X_s$.
    Let $\mathcal{B}_Q$ be the class of PWC functions in the form of Eq.~\eqref{eq:pwf} 
    and $\mathcal{P} = \mathcal{P}_1 \times \ldots \times \mathcal{P}_{|Q|}$, where each $\mathcal{P}_i$ is the set of feasible transitions.
    Then, $B^* \in \mathcal{B}_Q$ is a PWC-SBF if $B^*$ is a solution to \\
    $B^* = \arg\min_{B \in \mathcal{B}_{Q}} \; \max_{(p_i)_{i=1}^{|Q|} \in \mathcal{P}}
        \; \eta + N \beta $ 
        \;\;\; subject to, \;
    %
    \begin{align*}
        \forall i \in \{1,\ldots,Q\},
        \;\;
        b_i \geq 0, 
        \;\;
        \sum_{j = 1}^Q b_j \cdot p_{ij} + p_{i\unsafe} \leq  b_i + \beta_{i}, 
        \;\;
        0 \leq \beta_{i} \leq \beta,
    \end{align*}
    and for all $i: X_i \cap X_0 \neq \emptyset$, \;\; $b_i \leq \eta$.
    %
    Then, $B^*$ guarantees the safety probability in Eq.~\eqref{eq:prob}.
\end{theorem}
%
Work \cite{mazouz2024piecewise} introduces three methods for synthesizing PWC SBFs:
(i) an exact duality-based approach formulated as a zero-duality-gap Linear Program (LP),
(ii) an exact \emph{Counter-Example Guided Inductive Synthesis} (CEGIS) method that alternates between generating candidate barriers and counter-examples via two LPs, and
(iii) a \emph{gradient descent} (GD) method offering improved scalability and memory efficiency.
\tool implements all three methods for System~\eqref{eq:system} with linear and PWA inclusion dynamics. The toolbox internally computes the kernel bounds in Eq.~\eqref{eq:kernel bounds}.

\section{Overview of \textit{StochasticBarrier.jl} }
\label{sec:toolbox}
\begin{table}[t]
\caption{Overview of SBF synthesis methods implemented in \tool for different classes of dynamics.}
\vspace{-3mm}
\begin{center}
\begin{tabular}{@{}lll@{}}
\toprule
\textbf{Dynamics $f(x)$ \quad } & \textbf{SBF $B(x)$ \quad} & \textbf{Solver Method}\\ \midrule
Linear & SOS  & SDP     \\
       & PWC  & Dual LP, CEGIS LP, GD \\
       \hline
Polynomial  & SOS  & SDP        \\ 
        \hline
PWA Inclusion & SOS  & SDP   \\
(Eq.~\eqref{eqn:UpperandLowerBound})      
      & PWC  & Dual LP, CEGIS LP, GD \\
      \bottomrule
\end{tabular}
\end{center}
\label{tab:methods}
\end{table}

\tool~is a Julia toolbox for the synthesis of SBFs for System \eqref{eq:system}. The tool includes multiple independent synthesis methods with the following features:
\begin{itemize}
    \item Synthesis of SOS-SBFs for systems in the form of \eqref{eq:system} with linear, polynomial and PWA inclusion dynamics,
    \item 
    Computation of transition kernel bounds for System~\eqref{eq:system} with linear and PWA inclusion dynamics, 
    \item Synthesis of PWC-SBFs through 
    LP (dual and CEGIS) and GD optimization methods given the kernel bounds.
\end{itemize}

In Table~\ref{tab:methods}, the SBF methods are outlined for different classes of dynamics $f$. The \emph{solver method} refers to the convex optimization class. 
To instantiate an instance of SBF synthesis problem, the system model needs to be defined first, which consists of the dynamics and the relevant sets (safe, unsafe and initial).
In Section \ref{subsec:system}, we show how to set up these system related parameters. Then, the SBF synthesis method needs to be specified, as described in Section \ref{subsec:barrier}.

\subsection{System Setup}
\label{subsec:system}
\tool~dispatches and checks compatibility with the chosen synthesis algorithm based on the system type. We give examples about how to construct a linear, polynomial, and PWA inclusion system, and specify the safe, initial, and unsafe sets. The examples are for 2D-systems, but the user can easily set up an $n$-dimensional problem.

\subsubsection{Linear Dynamics Example}
\phantom{.}
With linear dynamics, both SOS and PWC classes of SBFs can be used.  
\paragraph{SOS-SBF.} To synthesize a SOS-SBF for System~\eqref{eq:system} with linear vector field $f(x) = Ax + b$, we first need to specify matrix $A$ and bias vector $b$.
The noise distribution is assumed to be zero-mean Gaussian $p_{\pw} =\mathcal{N}(0, \sigma^2 I)$ characterized by standard deviation $\sigma$. An example setup is shown below for a 2D-linear system.

\begin{lstlisting}
# System Linear: SOS
A = [0.95 0.0; 0.0 0.95], b = [0.0, 0.0], (*@{$\sigma$}@*) = [0.1, 0.1]
space = Hyperrectangle(low=[-1.0,-2.0], high=[1.0,2.0])
system = AdditiveGaussianLinearSystem(A, b, (*@{$\sigma$}@*), space)
\end{lstlisting} 
\vspace{-3mm}

The \juliafunction{Hyperrectangle} struct of \texttt{LazySets.jl}, a Julia package for calculus with convex sets \cite{lazysets21}, is used to define the state space. A struct is passed as a system object (\juliafile{AdditiveGaussianLinearSystem}) into \tool. 

\paragraph{PWC-SBF.} Using PWC-SBF requires an additional parameter to define a grid partitioning of the state space.  The tool uses this grid to define the domain of PWC function in Eq.~\eqref{eq:pwf} and compute the transition kernel bounds in Eq.~\eqref{eq:kernel bounds}. 
The grid is parameterized by $\epsilon \in \mathbb{R}^n$, where the $i$-th entry is the grid cell's half-width in the $i$-th dimension.
The \juliafile{generate\_partitions()} function takes the state space and $\epsilon$ as inputs, and returns the partitions. The system object \juliafile{AdditiveGaussianLinearSystem}, along with the state partitions, are parsed into \juliafile{transition\_probabilities()}, which is a function in \tool that computes the {transition kernel bounds}.
An example 2D-linear system setup is provided.
\begin{lstlisting}
# System Linear: PWC
A = [0.50 0.0; 0.0 0.50], b = [0.0, 0.0], (*@{$\sigma$}@*) = [0.1, 0.1]
space = Hyperrectangle(low=[-1.0,-1.0], high=[0.5,0.5])
# Half-width vector
(*@{$\epsilon$}@*) = [0.2, 0.2]
# Compute probability data
system = AdditiveGaussianLinearSystem(A, b, (*@{$\sigma$}@*))
partitions = generate_partitions(space, (*@{$\epsilon$}@*))
probabilities = transition_probabilities(system, partitions)
\end{lstlisting} 
\vspace{-3mm}

In addition, the toolbox supports saving and loading transition probability bounds to the Network Common Data Form (NetCDF) \cite{rew1990netcdf}. See~\cite{mazouz2026stochasticbarrier}[App. A] for more details on this feature.

\subsubsection{Polynomial Dynamics Example}
\phantom{.}
For polynomial systems, the tool only supports the SOS class of SBFs.
To set up, first define a vector of symbol variables $x$, using the \juliafunction{@polyvar} macro of \texttt{DynamicPolynomials.jl} \cite{legat2023multivariate}.
Then, the polynomial function $f$ is defined symbolically.  
The programmatic construct of \juliafile{AdditiveGaussianPolySystem} is passed as an object to the barrier synthesis method.
An example is provided below for a 2D system.
\begin{lstlisting}
# System Polynomial: SOS
@polyvar x[1:2]  # dim = 2
(*@{$\tau$}@*) = 0.1
f = [x[1] + (*@{$\tau$}@*) * x[2], x[2] + (*@{$\tau$}@*) * (-x[1] + (1 - x[1])^2)]
(*@{$\sigma$}@*) = [0.02, 0.02]
space = Hyperrectangle(low=[-6.0, -6.0], high=[6.0, 6.0])
system = AdditiveGaussianPolySystem(f, (*@{$\sigma$}@*), space)
\end{lstlisting} 
\vspace{-3mm}

\subsubsection{PWA Inclusion Dynamics Example}
\phantom{.}
Setting up the system model for PWA inclusion dynamics requires PWA bounds on dynamics $f$ of System~\eqref{eq:system}, as defined in Eq.~\eqref{eqn:UpperandLowerBound}. 
These bounds can be generated using Linear Bound Propagation \cite{zhang2018efficient}, and stored in a NetCDF file. These files can be loaded into
the tool, which offers the options of using SOS- or PWC-SBF methods.

\paragraph{SOS-SBF.} Given a NetCDF file that contains the PWA bounds, the setup for the SOS-SBF option is as follows. First, the system flag and filename need to be defined, which together specify the path to the PWA bounds. Then, the 
\juliafile{load\_dynamics()} function reads this data. The bounds dataset, along with the standard deviation $\sigma$, are passed into the  \juliafile{AdditiveGaussianUncertainPWASystem} struct. See the example below for a 2D pendulum model.
\begin{lstlisting}
# System
system_flag = "pendulum"
filename    = "path/to/$system_flag/data.nc"
dataset     = open_dataset(filename)
data        = load_dynamics(dataset)
(*@{$\sigma$}@*) = [0.01, 0.01]
system = AdditiveGaussianUncertainPWASystem(data, (*@{$\sigma$}@*))
\end{lstlisting}
\vspace{-3mm}

\paragraph{PWC-SBF.} 
For the PWC-SBF option, the file and system specification are similar. In addition, the transition probability data must be computed using  \juliafile{transition\_probabilities()}.

\begin{lstlisting}
# Compute probability data
probabilities = transition_probabilities(system)
\end{lstlisting}
\vspace{-3mm}

\subsubsection{Specifying Initial and Unsafe Sets}
\phantom{} 
To complete the system model setup, the initial set and unsafe region(s) need to be defined. This is done by using the \juliafunction{Hyperrectangle} struct. 
In the toolbox, the region $\reals^{n} \setminus X$ is considered unsafe by default. The user can further declare unsafe regions in the domain. If more than one unsafe region is defined, the \juliafunction{UnionSet} function of 
\texttt{LazySets.jl} can be used to take their union, as shown in the example below. 
\begin{lstlisting}
initial = Hyperrectangle(low=[-0.2, -0.3], high=[0.0, 0.1])
unsafe1 = Hyperrectangle(low=[-0.55, 0.4], high=[-0.45, 0.6])
unsafe2 = Hyperrectangle(low=[0.45, 0.4], high=[0.55, 0.6])
unsafe_regions = UnionSet(unsafe1, unsafe2)
\end{lstlisting} 
\vspace{-3mm}

If domain $X$ does not contain any unsafe regions, the variable \texttt{unsafe\_regions}
must be declared empty:  
\juliafunction{EmptySet(dim)}, where \textit{dim} is the dimension of the system. With that, the system model setup is concluded, and one can call on various SBF synthesis methods, as explained below.


\subsection{SBF Synthesis} 
\label{subsec:barrier}
All methods for synthesizing SBFs are called under
\juliafile{synthesize\textunderscore barrier(alg, system, initial\textunderscore region, unsafe\textunderscore region; time\textunderscore horizon=N)}, which is designed with a unified interface of input parameters.
Julia's multiple dispatch will select the appropriate method based on the \juliafile{alg} parameter, and each algorithm returns a result object, which contains, in addition to the barrier, the resulting safety certificate and computation time. The barrier type is determined by the algorithm.
Specifying the time horizon is optional, with the default value set to $N = 1$.

\subsubsection{SOS-SBF}
\phantom{.} 
To synthesize an SOS-SBF, \juliafile{synthesize\textunderscore barrier} is called with input \juliafile{SumOfSquaresAlgorithm()} as the first parameter. The second parameter, the system model object defined in Section \ref{subsec:system}, can be any of the three system types.

\begin{lstlisting}
N = 10
# Optimize: baseline 1 (sos approach)
res = synthesize_barrier(SumOfSquaresAlgorithm(), system,
    initial_region, unsafe_region; time_horizon = N)

# Result
barr = barrier(res)
(*@{$\eta$}@*), (*@{$\beta$}@*), ps = eta(res), beta(res), psafe(res, N)
computation_time = total_time(res)
\end{lstlisting} 
\vspace{-3mm}

The optimization returns a struct of values, which include the SOS-SBF $B$, $\eta$, and $\beta$ values, and the computation time. 
The algorithm \juliafile{SumOfSquaresAlgorithm()} accepts keyword arguments to configure, e.g. the underlying SDP solver and the degree of the SOS barrier. Below we list the default values and how to construct the object for customization. 
\begin{lstlisting}
alg = SumOfSquaresAlgorithm(;
    barrier_degree  = 4
    lagrange_degree = 2
    sdp_solver = Mosek.Optimizer)
\end{lstlisting}
\vspace{-3mm}

Setting \juliafile{SumOfSquaresAlgorithm(barrier\textunderscore degree = 30)}, for example, overrides the barrier degree.
The default SDP solver for SOS optimization is  Mosek \cite{aps2019mosek}. 

\subsubsection{PWC-SBF}
For PWC-SBFs, \juliafile{synthesize\textunderscore barrier} is called with first parameter 
\juliafile{DualAlgorithm()} for the dual LP approach, \juliafile{CEGISAlgorithm()} for the CEGIS approach, or \juliafile{GradientDescentAlgorithm()} for the GD approach, respectively. The second parameter should be the result from \juliafile{transition\textunderscore probabilities}.


\begin{minipage}{\linewidth}
\begin{lstlisting}
N = 10
# Optimize: method 1 (dual approach)
dual = synthesize_barrier(DualAlgorithm(), probabilities,   
    initial_region, obstacle_region, time_horizon = N)

# Optimize: method 2 (CEGIS approach)
cegis = synthesize_barrier(CEGISAlgorithm(), probabilities,
    initial_region, obstacle_region; time_horizon = N)

# Optimize: method 3 (GD approach)
gd = synthesize_barrier(GradientDescentAlgorithm(), probabilities,
    initial_region, obstacle_region; time_horizon = N)
\end{lstlisting}
\end{minipage}
\vspace{-3mm}

Similar to SOS, each algorithm returns a result struct, which, in addition to the PWC-SBF $B$, values for $\eta$, $\beta$ and the computation time, also includes $\beta_i$, which is an upper bound on Eq. \eqref{eq:barrier_expectation} in each region $i$.
Next, we highlight the core properties of each algorithm.


\paragraph{Linear Programming.}
As outlined in Section \ref{subsec:pwcmethods}, amongst the LP classes, there is the dual LP method, and the CEGIS method which uses inner and outer LP optimizations. 
The dual LP method only has the solver as a configurable parameter.
\begin{lstlisting}
alg = DualAlgorithm(;
    linear_solver = HiGHS.Optimizer)
\end{lstlisting}
\vspace{-3mm}
The default LP solver across the toolbox is HiGHS solver \cite{huangfu2018parallelizing}, but is customizable to any LP solver supported by JuMP, Julia Mathematical Programming \cite{Lubin2023}.

The \juliafile{CEGISAlgorithm} can additionally be configured with the number of iterations.
\begin{lstlisting}
alg = CEGISAlgorithm(;
    linear_solver = HiGHS.Optimizer
    num_iterations = 10
    distribution_guided = true)
\end{lstlisting}
\vspace{-3mm}

The default is the distribution guided approach, where in each iteration one LP generates a candidate PWC-SBF that optimizes for the  probability of safety, given a set of finite feasible distributions. Then, the second LP generates distribution witnesses that violate the safety probability guarantee of the candidate PWC-SBF. By default, it runs for a fixed number of iterations given by  \textit{num\_iterations} (default = 10), or it can continue adaptively until no further counterexamples exist, guaranteeing finite-time convergence~\cite{mazouz2024piecewise}.

\paragraph{Projected Gradient Descent.}
The GD method solves the same problem as the LP approaches, utilizing the minimax formulation to define a convex loss function.
The projected gradient descent method is subject to hyperparameters. 

\begin{lstlisting}
alg = GradientDescentAlgorithm(;
    num_iterations = 10000
    initial_lr = 1e-2
    decay = 0.9999
    momentum = 0.9)
\end{lstlisting} 
\vspace{-3mm}

The struct \juliafile{GradientDescentAlgorithm} defines key parameters for running GD. The \textit{num\_iterations} parameter controls the maximum number of iterations (default: 10,000). The \textit{initial\_lr} parameter sets the initial learning rate for the algorithm (default: 0.01), determining the size of each update step. The \textit{decay} parameter (default: 0.9999) reduces the learning rate over time to fine-tune convergence. Finally, \textit{momentum} (default: 0.9) helps to accelerate convergence by incorporating a portion of the previous gradient step into the current step. 
Importantly, at termination, the values for $\eta$, $\beta$ are guaranteed to be sound with respect to the returned barrier, although not absolutely optimal in the sense of Theorem \ref{th:pwa}.

\tool is also equipped with an independent interface to easily run benchmarks. See~\cite{mazouz2026stochasticbarrier}[App. A] for details.

\begin{table}[b]
\caption{Linear, polynomial and PWA inclusion stochastic dynamic systems 
for benchmarking in \tool.
\vspace{-3mm}
}
\begin{center}
\begin{tabular}{llc}
\toprule
\textbf{Dynamics}   &\textbf{Name }                  & \textbf{Dimension} \\ \hline
Linear        & Contraction Map 1 \& 2        & 2         \\
              & Two Tank \cite{ramos2007mathematical}      & 2         \\
                & Room Temperature \cite{wooding2024protect}      & 3         \\
              & Quadrotor \cite{mazouz2024piecewise}             & 6         \\
\hline
Polynomial    & Thermostat \cite{nejati2021compositional}           & 1         \\
              & Van der Pol Oscillator \cite{abate2020arch} & 2         \\
             \hline
PWA Inclusion & Pendulum NNDM \cite{mazouz2022safety}         & 2         \\
              & Unicycle \cite{{mazouz2024piecewise}}               & 4         \\ 
\bottomrule 
\end{tabular}
\end{center}
\label{tab:benchmarks}
\end{table}

\begin{figure*}[t!]
    \centering
    \begin{minipage}{0.33\textwidth}
        \centering
        \includegraphics[width=1.0\textwidth, trim = 1cm 6.4cm 2cm 6cm, clip]{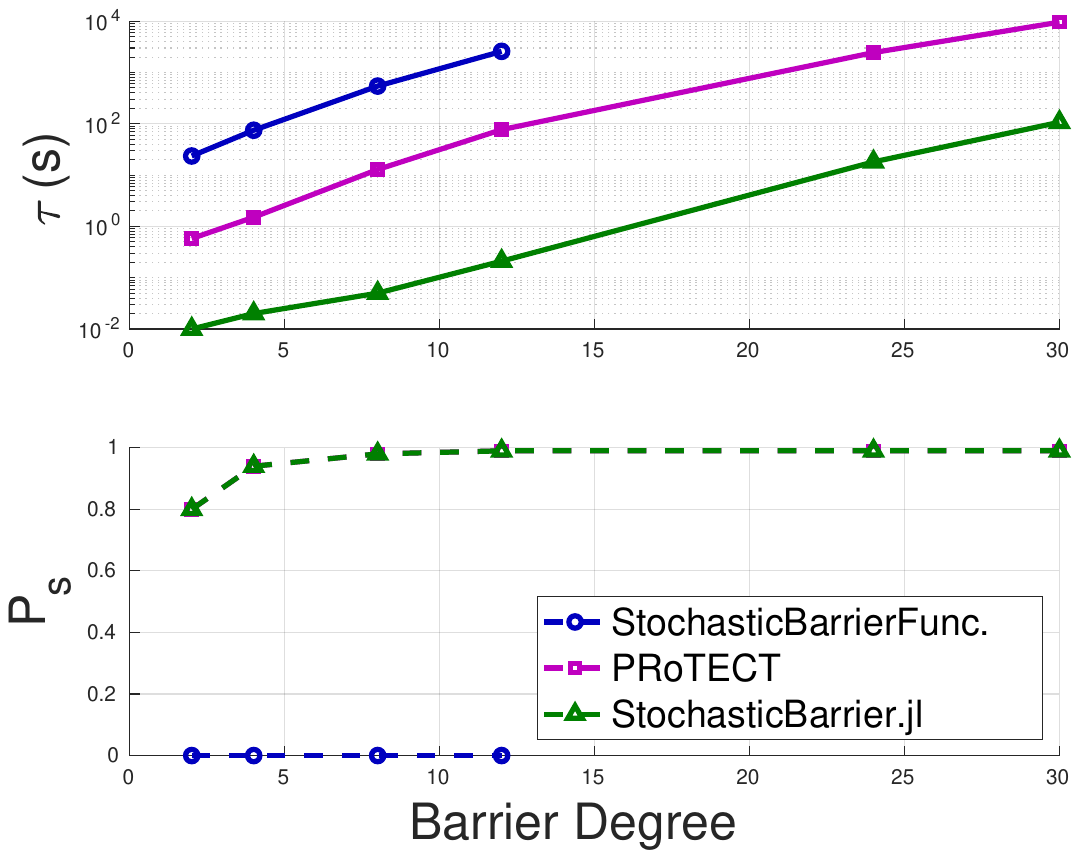}
                \\(a) Contraction Map
    \end{minipage}%
    \begin{minipage}{0.33\textwidth}
        \centering
        \includegraphics[width=1.0\textwidth, trim = 1cm 6.4cm 2cm 6cm, clip]{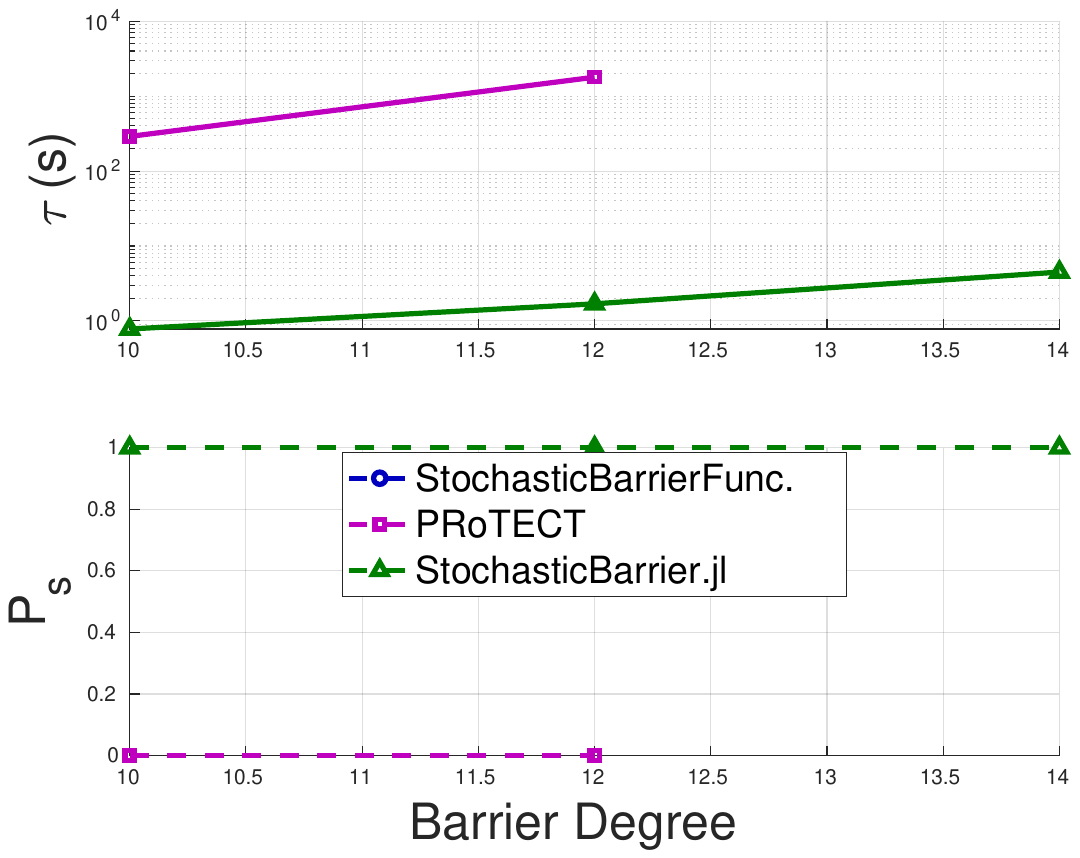}
                \\(b) Two Tank
    \end{minipage}
        \begin{minipage}{0.33\textwidth}
        \centering
        \includegraphics[width=1.0\textwidth, trim = 1cm 6.4cm 2cm 6cm, clip]{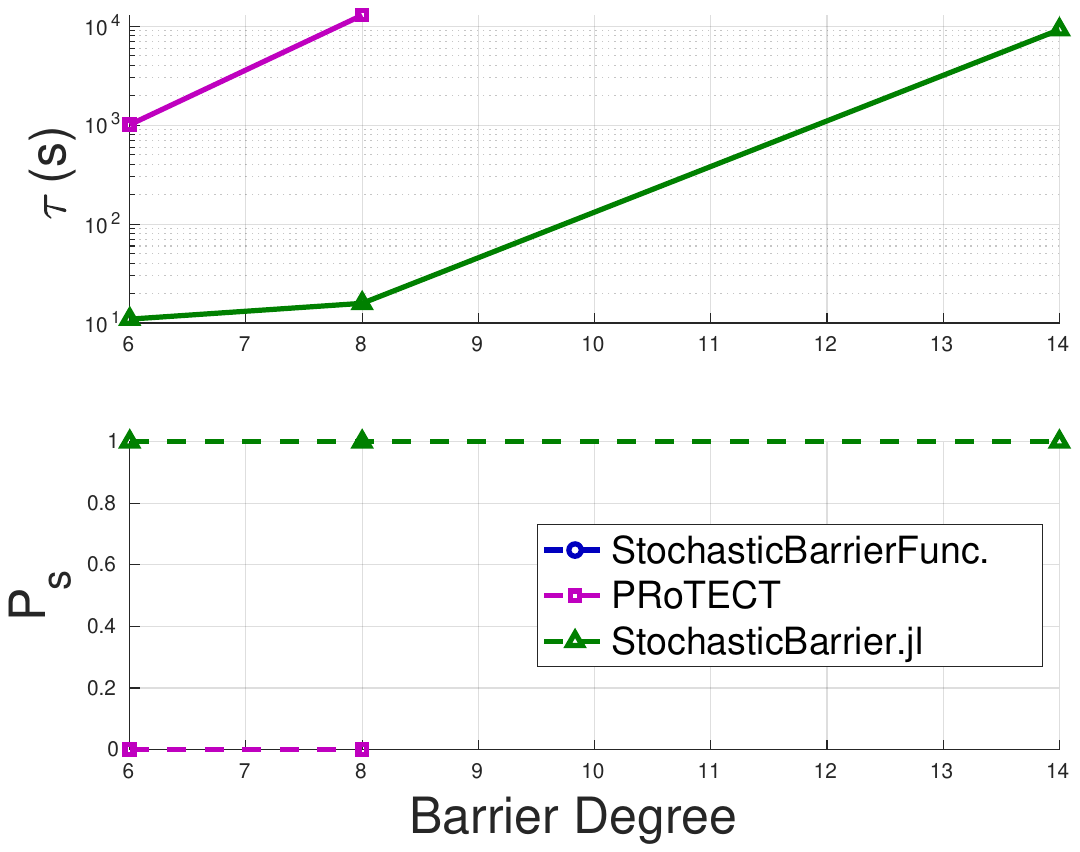}
        \\ (c) Room Temperature
    \end{minipage}
    \begin{minipage}{0.33\textwidth}
        \centering
        \includegraphics[width=1.0\textwidth, trim = 1cm 6.4cm 2cm 6cm, clip]{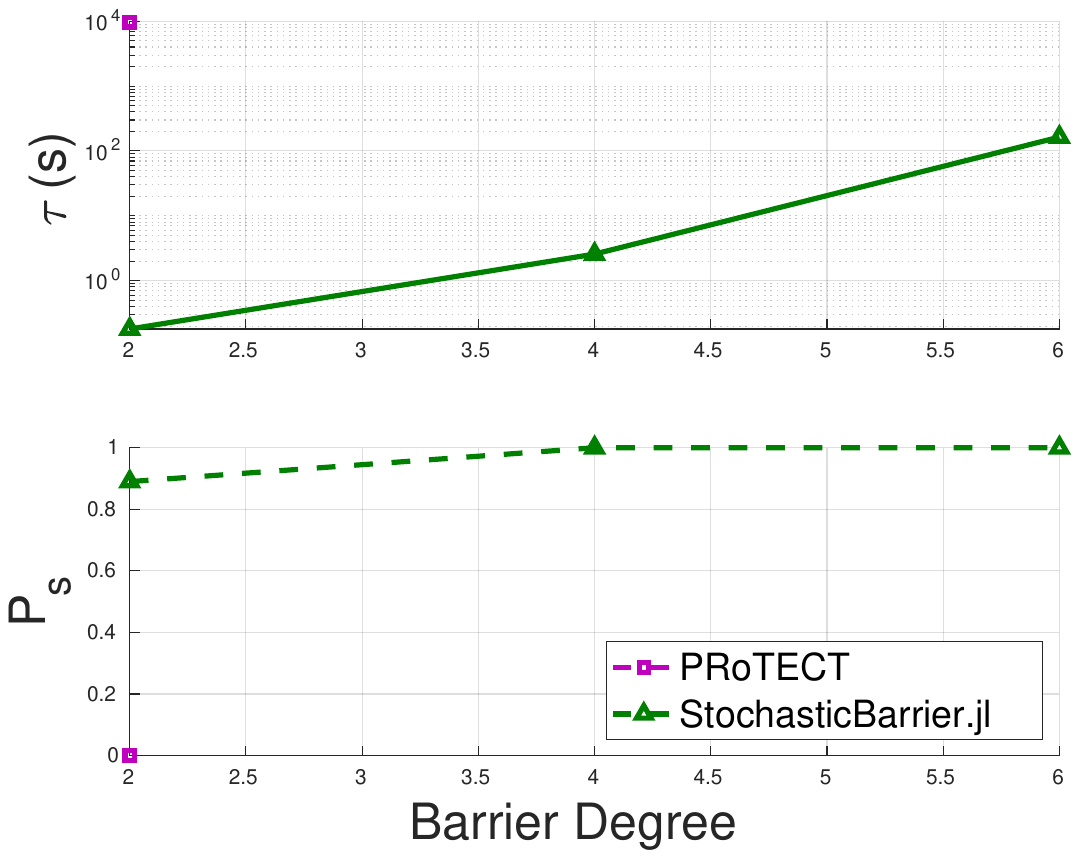}
        \\ (d) Quadrotor
    \end{minipage} 
\vspace{3mm}
        \begin{minipage}{0.33\textwidth}
        \centering
        \includegraphics[width=1.0\textwidth, trim = 1cm 6.4cm 2cm 6cm, clip]{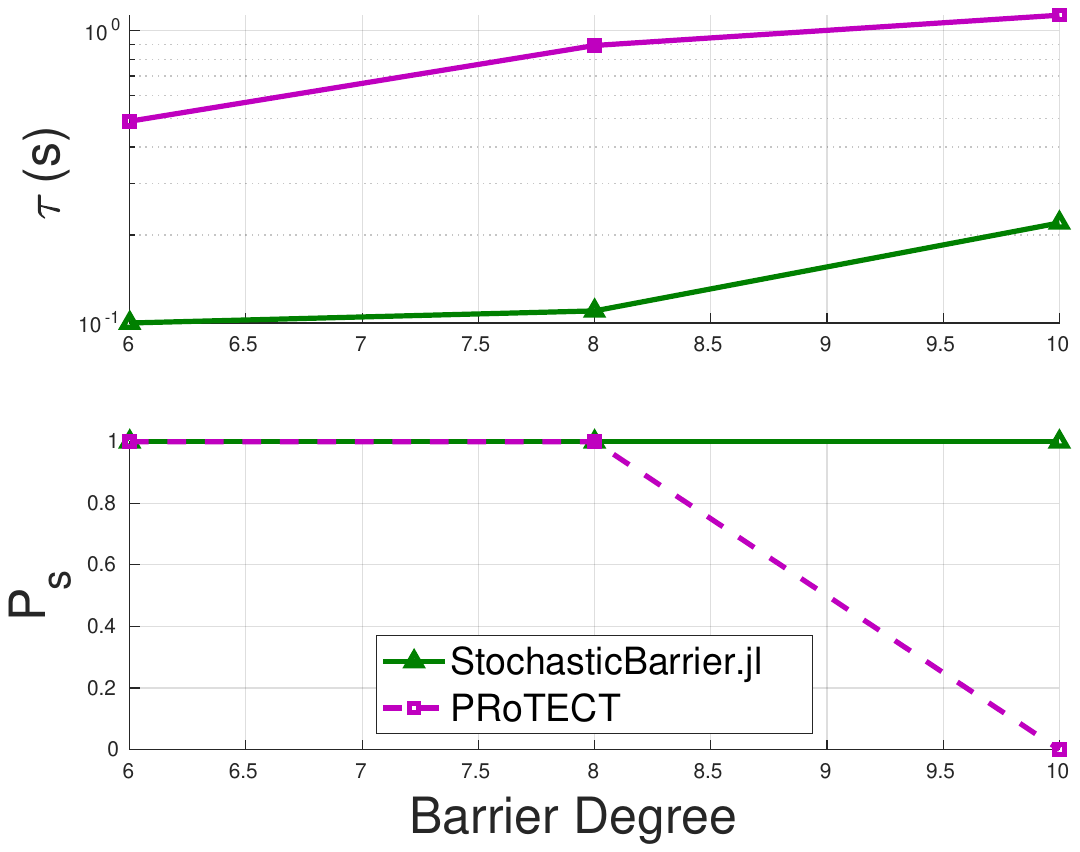}
        \\(e) Thermostat
    \end{minipage}%
    \begin{minipage}{0.33\textwidth}
        \centering
        \includegraphics[width=1.0\textwidth, trim = 1cm 6.4cm 2cm 6cm, clip]{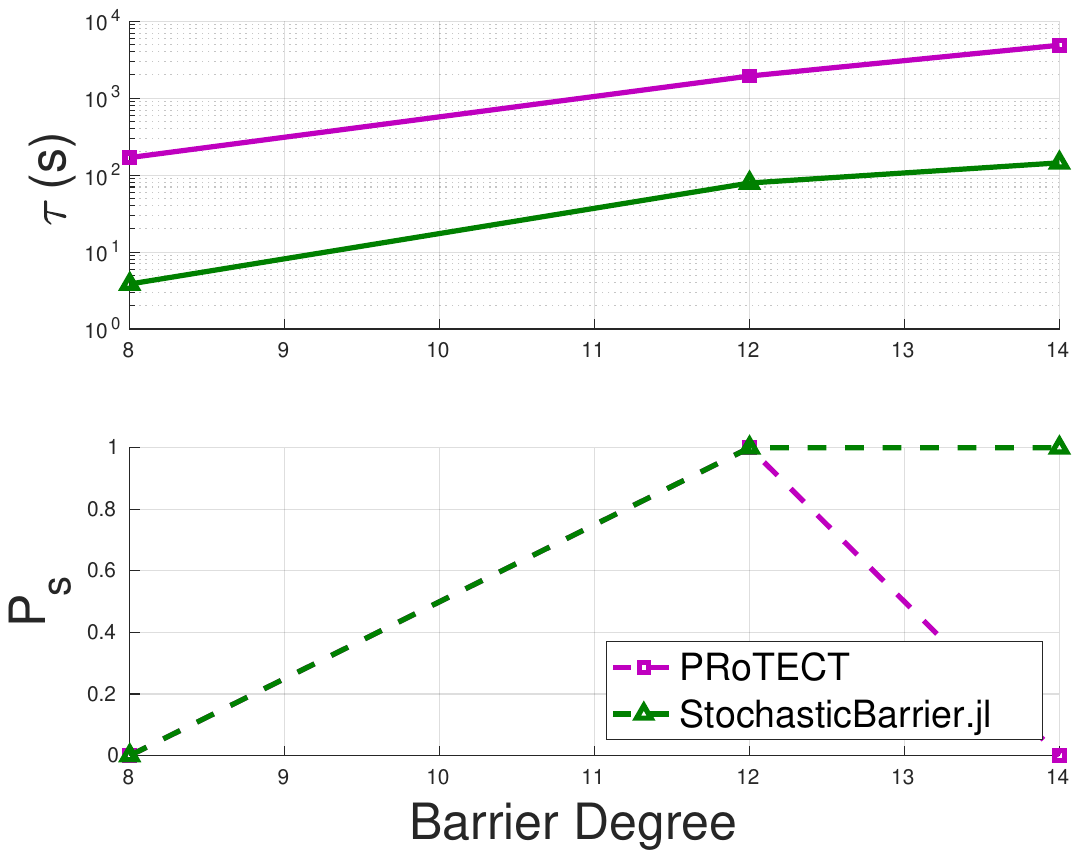}
        \\(f) Oscillator
    \end{minipage}
    \caption{
    Polynomial systems benchmarking of \texttt{StochasticBarrier.jl} \emph{vs} StochasticBarrierFunctions \cite{SANTOYO2021109439} and PRoTECT \cite{wooding2024protect}. Plots depict optimization time $\tau (s)$ and probability of safety $P_s$. If data points are not plotted, the algorithm ran out of memory.
    }
    \label{fig:SOSbench}
\end{figure*}

\section{Benchmark Studies}
\label{sec:studies}
In this section, we present benchmark results 
for various stochastic systems with linear, polynomial and PWA inclusion dynamics. 
We performed
two main comparison studies.
First, the SOS-SBF algorithm in \tool is compared  against the state-of-the-art SOS toolboxes:
\texttt{StochasticBarrierFunctions} \cite{SANTOYO2021109439} (MATLAB) and
\texttt{PRoTECT} \cite{wooding2024protect} (Python). 
The second study evaluates all the methods (SOS, DUAL-LP, CEGIS-LP, and GD) in \tool and compares their performance.
All experiments are performed on a computer with a 3.9 GHz 8-Core CPU and 128 GB of memory. All methods use Mosek as the underlying SDP solver.
The systems considered are summarized in Table~\ref{tab:benchmarks} and 
 details of the systems and their setups are provided in~\cite{mazouz2026stochasticbarrier}[App. B.]
For all benchmarks, time horizon $N=10$ is used. The tool is available at  \url{https://github.com/aria-systems-group/StochasticBarrier.jl}.

\textbf{Comparison Against Other SOS-SBF Tools. }
Fig.~\ref{fig:SOSbench} shows performance of the SOS-SBF algorithm in \tool against state-of-the-art \texttt{StochasticBarrierFunc.} \cite{SANTOYO2021109439} and \texttt{PRoTECT} \cite{wooding2024protect}. A detailed table of results can be found in~\cite{mazouz2026stochasticbarrier}[App. D]. Note that as \texttt{StochasticBarrierFunc.} only supports linear systems, and \texttt{PRoTECT}
only linear and polynomial systems,
we could only compare the tools with systems of this dynamics class. 
We also note that the results from \texttt{PRoTECT} are random, and we present average results over 10 runs. We also note that $P_s$ defined in Eq.~\eqref{eq:prob} is computed differently in \texttt{PRoTECT}; we adjust it here to match our setting. This paper solely focuses on optimization-based SOS methods, i.e., disabling the objective is fundamentally different and not considered. 

Across all benchmarks, \tool outperforms both \texttt{PRoTECT} and \texttt{StochasticBarrierFunc.} in terms of computation time and scalability, and tightness of safety probability.
\texttt{StochasticBarrierFunc.} fails to produce non-trivial results and cannot handle higher-degree SOS-SBFs (beyond degree 12), often yielding a trivial $P_s \ge 0$ or exceeding available memory.
\texttt{PRoTECT} performs better but remains significantly slower (by factors ranging from $5\times$ to $1000\times$) and often returns trivial safety bounds for complex systems.
In contrast, \tool, scales efficiently to high-degree SOS-SBFs (up to 30), achieves non-trivial or high safety probabilities, and maintains orders-of-magnitude faster computation across all tested models (contraction map 1, two tank, room temperature, quadrotor, thermostat, and oscillator).

\begin{figure*}[t!]
    \centering
    \begin{minipage}{0.33\textwidth}
        \centering
        \includegraphics[width=1.0\textwidth, trim = 1cm 6.5cm 2cm 6cm, clip]{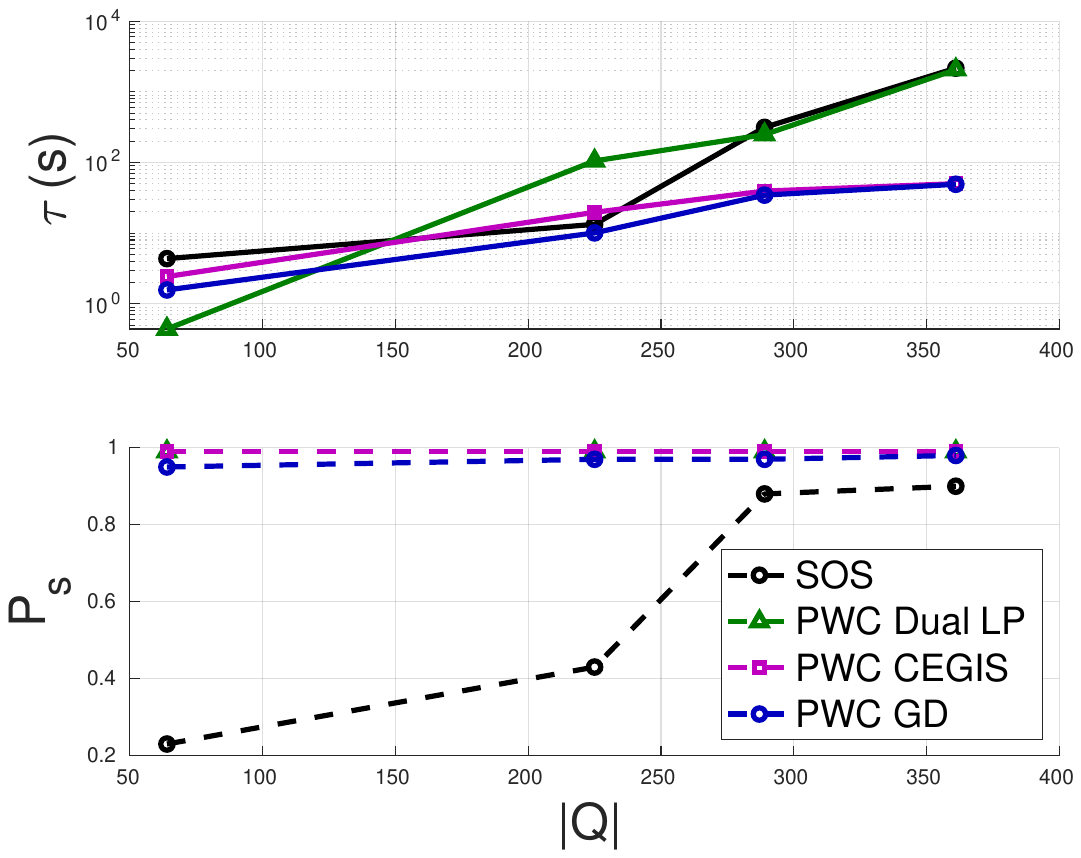}
        \\ (a) Contraction Map
    \end{minipage}%
    \begin{minipage}{0.33\textwidth}
        \centering
        \includegraphics[width=1.0\textwidth, trim = 1cm 6.5cm 2cm 6cm, clip]{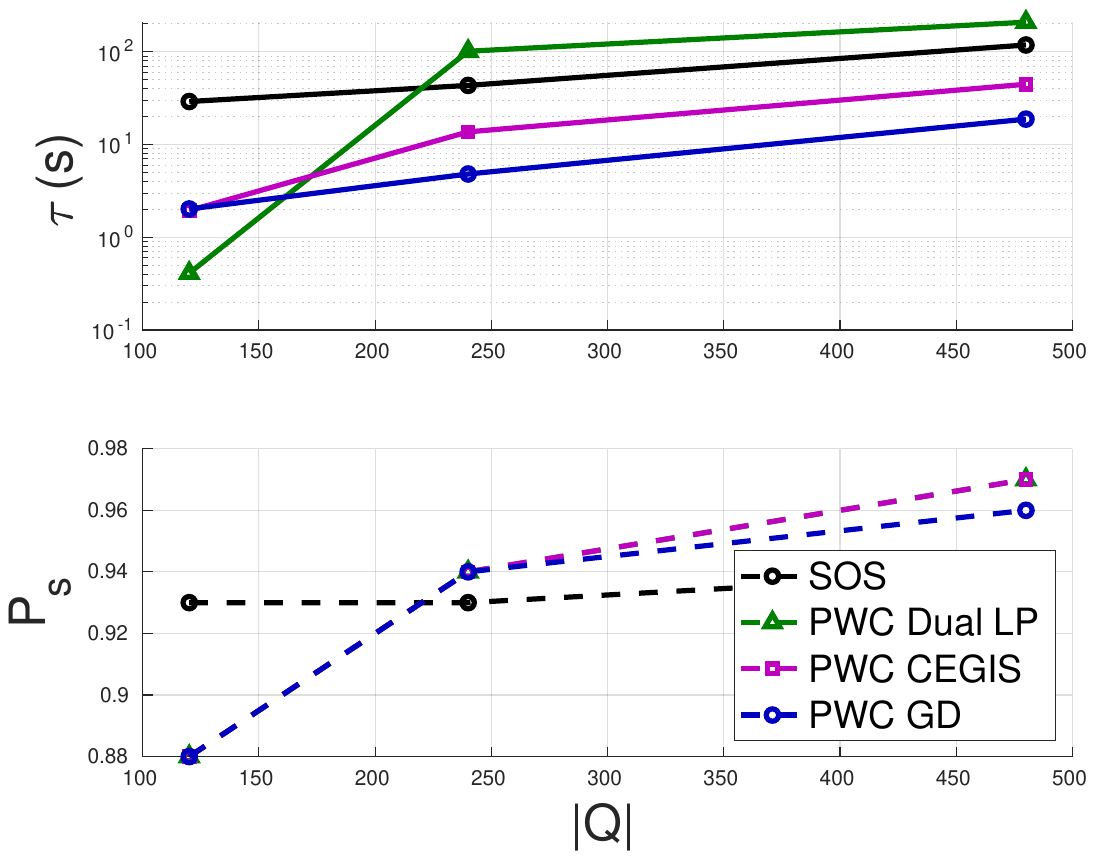}
        \\(b) Pendulum
    \end{minipage}
    \begin{minipage}{0.33\textwidth}
        \centering
        \includegraphics[width=1.0\textwidth, trim = 1cm 6.5cm 2cm 6cm, clip]{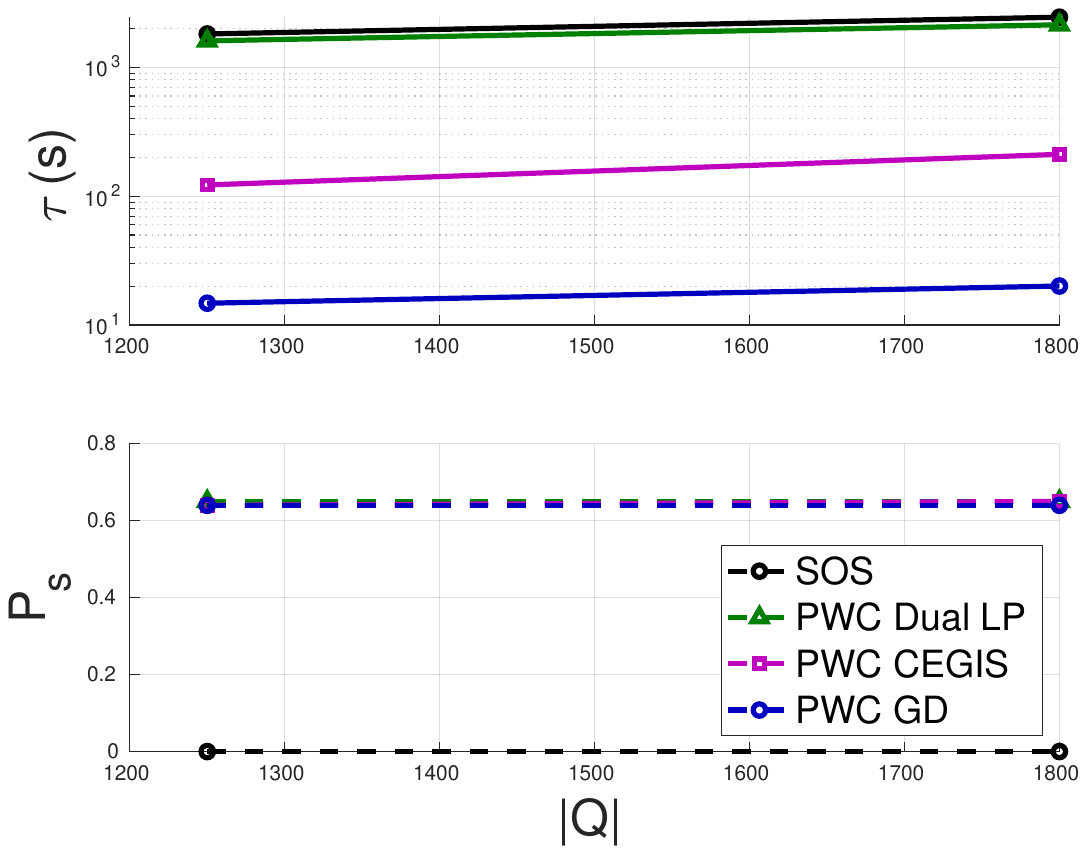}
        \\(c) Unicycle
    \end{minipage} \\
    \caption{Benchmark plots for the three PWC-SBF methods. For the pendulum and unicycle, the SOS results are also depicted for comparison. Plots depict total optimization time $\tau (s)$ and probability of safety $P_s$.}
    \label{fig:PWCbench}
\end{figure*}

\textbf{Comparisons of \tool engines. }
Benchmark results for the three PWC-SBF methods implemented in \tool are summarized in Fig.~\ref{fig:PWCbench}, with detailed results presented in~\cite{mazouz2026stochasticbarrier}[App. D]. The results show that PWC-SBFs consistently outperform SOS-SBFs, particularly for the contraction map and hybrid nonlinear systems. For the linear contraction map, the GD method achieves the fastest computation, especially as the number of partitions increases, while Dual LP and CEGIS yield tighter safety probability bounds that converge with higher partition counts. For PWA inclusion systems, PWC-SBFs are up to two orders of magnitude faster, and for the unicycle model, they provide non-trivial safety bounds where SOS fails. As expected, the Dual LP and CEGIS methods deliver the tightest bounds at the cost of higher computation time.

\section{Conclusion}
\label{sec:conclusion}

We introduced \tool, a Julia toolbox for generating barrier certificates and estimating safety probabilities of discrete-time stochastic dynamical systems. Benchmark results show that \tool is orders of magnitude faster and provides tighter safety bounds than existing MATLAB and Python implementations, with superior scalability to higher-dimensional systems and higher-degree SOS barriers. The toolbox also supports piecewise constant barriers, which consistently outperform SOS formulations. Current limitations include its focus on additive Gaussian noise. Future extensions will address non-Gaussian and non-additive noise. 


\bibliographystyle{splncs04}
\bibliography{cite}


\end{document}